# Observation of Majorana fermions with spin selective Andreev reflection in the vortex of topological superconductor

Hao-Hua Sun[1†], Kai-Wen Zhang[2†], Lun-Hui Hu[3,4], Chuang Li[3,4], Guan-Yong Wang[1], Hai-Yang Ma[1], Zhu-An Xu[3,4], Chun-Lei Gao[1,4], Dan-Dan Guan[1,4], Yao-Yi Li[1,4], Canhua Liu[1,4], Dong Qian[1,4], Yi Zhou[3,4], Liang Fu[5], Shao-Chun Li[2,4*], Fu-Chun Zhang[3,4*] and Jin-Feng Jia[1,4*]

[1]Key Laboratory of Artificial Structures and Quantum Control (Ministry of Education), Department of Physics and Astronomy, Shanghai Jiao Tong University, Shanghai 200240, China
[2]National Laboratory of Solid State Microstructures and School of Physics, Nanjing University, Nanjing 210093, China
[3]Department of Physics, Zhejiang University, Hangzhou 310027, Zhejiang, China
[4]Collaborative Innovation Center of Advanced Microstructures, Nanjing 210093, China
Department of Physics, Massachuset Institute of Technology, MA, USA

[†] These authors contributed equally;

**Majorana fermion (MF) whose antiparticle is itself has been predicted in condensed matter systems. Signatures of the MFs have been reported as zero energy modes in various systems. More definitive evidences associated with MF's novel properties are highly desired to verify the existence of the MF. Very recently, theory has predicted MFs to induce spin selective Andreev reflection (SSAR), a novel magnetic property which can be used to detect the MFs. Here we report the first observation of the SSAR from MFs inside vortices in $Bi_2Te_3/NbSe_2$ hetero-structure, in which topological superconductivity was previously established. By using spin-polarized scanning tunneling**

**microscopy/spectroscopy (STM/STS), we show that the zero-bias peak of the tunneling differential conductance at the vortex center is substantially higher when the tip polarization and the external magnetic field are parallel than anti-parallel to each other. Such strong spin dependence of the tunneling is absent away from the vortex center, or in a conventional superconductor. The observed spin dependent tunneling effect is a direct evidence for the SSAR from MFs, fully consistent with theoretical analyses. Our work provides definitive evidences of MFs and will stimulate the MFs research on their novel physical properties, hence a step towards their statistics and application in quantum computing.**

Majorana fermion (MF) is a special type of fermion whose anti-particle is itself[1]. MF was initially proposed in elementary particle physics, and the recent effort in searching such a genuine particle focuses on the neutrino-less double beta decay experiment[2]. MF may emerge as novel excitation in some condensed matter systems. They obey non-Abelian statistics and may be used as robust building blocks in quantum computing [3,4].

Chiral *p*-wave superconductors[5] (SCs) and $v=5/2$ fractional quantum Hall system[6] are possible candidates to host MF[6]. Fu and Kane proposed the existence of MFs at the interface of topological insulator (TI) and *s*-wave SC[7]. In recent years, a number of proposals have been explored to detect MFs, including the experiments in one-dimensional (1D) Rashba semiconducting wire[8-11] and 1D Fe atom chains on Pb[12]. Topological superconductivity may be induced on surfaces of 3D TIs such as $Bi_2Se_3$, and $Bi_2Te_3$ via proximity effect[13-17], and the localized MF at the vortex core has been studied by using STM/STS[13-15]. However, there are quasi-particle states inside the vortex core, whose lowest energies cannot be distinguished from the zero modes within the present STM resolution. This introduces great difficulty in detection of the MFs by using normal STM/STS with a non-polarized tip. He et al. have recently proposed a novel property of MF in 1D system[18], where the Andreev reflection (AR) is spin selective. They showed that MF induces selective equal spin AR, in which incoming electrons with certain spin polarization in the lead are reflected as counter-propagating holes with the same spin[7]. The proposed spin selective AR of MF in 1D wire can be generalized to 2D topological SC. In the latter case, the spin of the MF in the vortex core has a spatial distribution[15,16]. Nevertheless, its spin is fully polarized along the external magnetic field at the center of the vortex core $r=0$, with r the lateral distance of the tip from the vortex center, see Fig. 3A below. Therefore, the AR at $r=0$ is expected to be spin selective, and can be probed in spin polarized STM/STS[18-20].

A unique advantage to probe spin selective AR in the topological SC is that a small magnetic field is sufficient to induce vortices, hence the spin-polarized MF in the vortex core. On the other hand, the spin polarization of surface states and finite-energy quasiparticle states is still negligibly small, as we can see in Fig. 3B below. In other systems, it is required to apply a sufficiently large magnetic field to make the system topological in the first place, in which case the bands are already spin-polarized. It will then be impossible to attribute spin-dependent zero-bias conductance to MF since both MF (if present) and finite-energy mid gap states would be spin-polarized.

In Fig. 1, we illustrate SSAR induced by the MF at the vortex center of an interface of TI and SC. The tunneling conductance consists of two parts,

$$\frac{dI(r,E;\hat{B},\hat{M})}{dV} = \frac{dI(r,E;\hat{B},\hat{M})}{dV}\Big|_n + \frac{dI(r,E;\hat{B},\hat{M})}{dV}\Big|_A \quad (1)$$

Where the first term is the contribution from the normal tunneling, and the second term from the AR. E is the energy, $\hat{B}$ and $\hat{M}$ are the orientations of the external magnetic field B and spin polarization M, respectively. Note that the SSAR is most profound at $r = 0$, where the superconducting order parameter vanishes, hence the AR is purely induced via the MF. Hereafter we shall focus on the discussion for low energy spectra at $r = 0$ unless explicitly specified otherwise. dI/dV|n is proportional to the local density of states, which is independent of spin polarization at $r = 0$ and $E = 0$ within the energy resolution of about 0.1 meV (see model calculation part). Therefore, we expect the first term in Eq. (1) is independent of spin polarization, and the difference of spin-dependent conductance probes the SSAR.

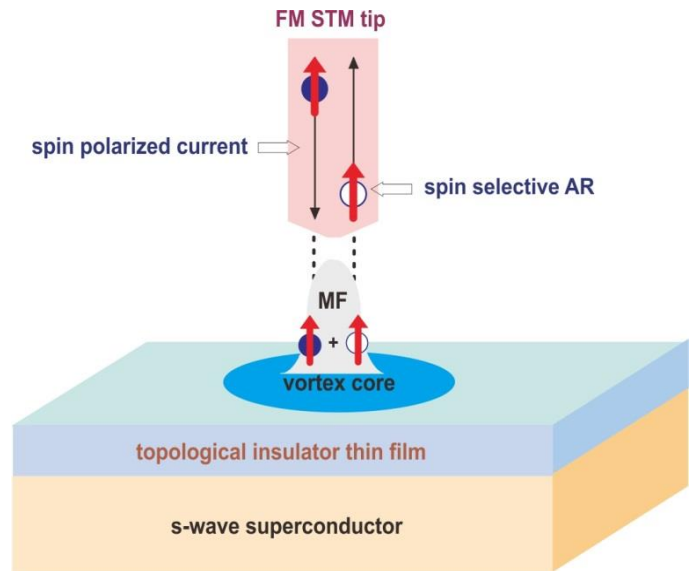

**Figure 1 | Illustration of the spin selective Andreev reflection (SSAR)**. SSAR in spin polarized (M↑) STM/STS on a vortex center r=0 in an interface of a topological insulator and s-wave superconductor is shown. An incoming spin-up electron is reflected as an outgoing spin-up hole induced by Majorana fermion with spin up at r=0.

In this work, we report the observation of MF via spin polarized-STM/STS measurements. Ferromagnetic Fe/W tips are applied to probe vortex core states in 5 quintuple layers (QL) $Bi_2Te_3$/$NbSe_2$ hetero-structure, where the topological superconductivity has been established[14,15,17]. The intensity of the zero bias peak (ZBP) at the vortex center is observed to be dependent on the magnetic polarization M of the applied tip, and it is 14% higher for M parallel than anti-parallel to the applied magnetic field B. We attribute the spin dependent tunneling to the SSAR, a special novel property of the MF in SC. The experiment observation is in good agreement with a model calculation.

Abrikosov vortex has recently been observed on TI/SC hetero-structure[14]. At the center of vortex core, a typical ZBP can be seen. Because of the existence of MF on

TI/SC, the splitting of the spectra of the core states from the center shows a Y-shape, deviating from V-shape in the usual SC[15,20,21]. Fig. 2A shows the mapping of one 5QL $Bi_2Te_3$/$NbSe_2$ sample. The out-of-plane magnetic field applied on sample was B=0.1 T. The size of the vortex was approximately 80 nm.

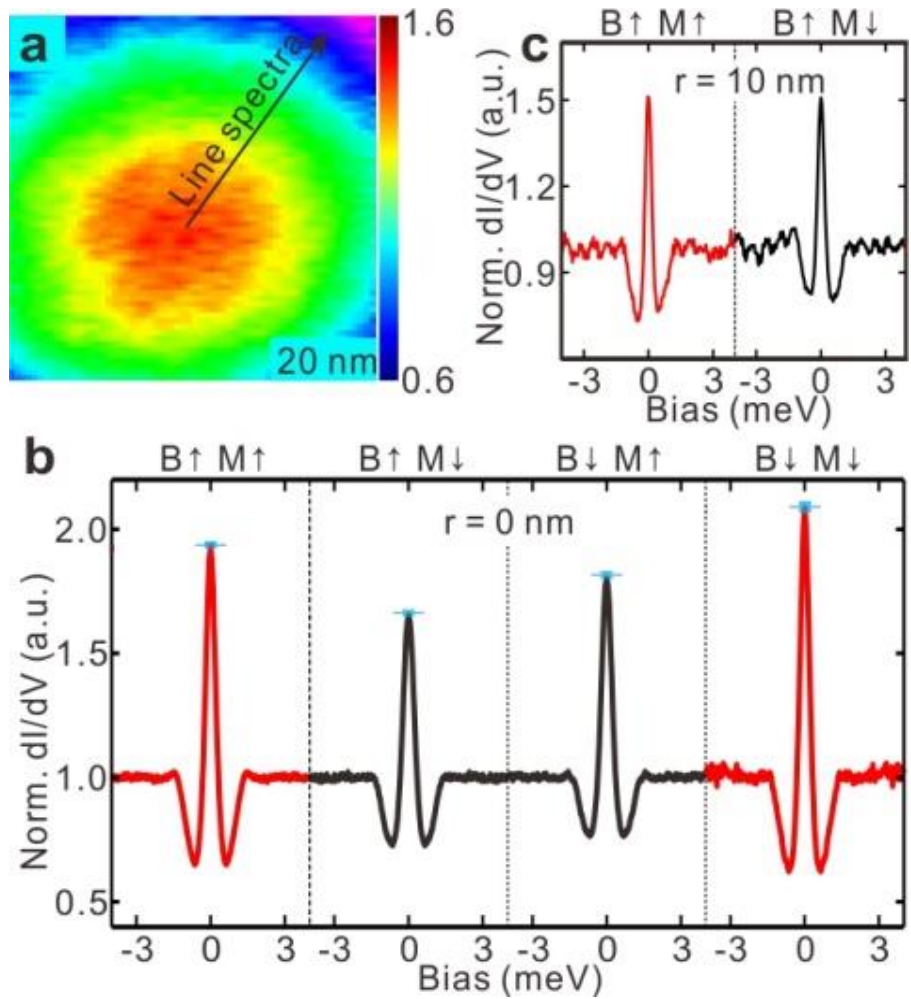

**Figure 2 | Zero bias dI/dV at a vortex core on 5QL $Bi_2Te_3$/$NbSe_2$. a.** Zero bias dI/dV mapping of a vortex at 0.1 T with spin non-polarized tip on topological superconductor 5QL $Bi_2Te_3$/$NbSe_2$. **b.** dI/dV at the vortex center measured with a fully spin polarized tip. Red curves are for tip polarization M parallel to magnetic field B, and black curves are for M anti-parallel to B. In the measurements, B=0.1 T and temperature T=30 mK. The blue lateral lines give the average values of the intensities in multi measurements, the vertical bars are the standard error bars. The intensity of the conductance with M parallel to B is about 14% higher than that with M anti-parallel to B. **c.** dI/dV at 10nm away from the center of a vortex measured with a fully spin polarized tip, where the tunneling is found independent of the spin polarization.

Fe coated W tips were adopted to study the ZBP at the vortex core of the TI/SC 5 QL sample. Prior to the deposition of Fe atoms, the W tip was annealed at a temperature over 2200 K to remove oxide layers. A magnetic field of 2.0 T perpendicular to the plane was applied on and then removed gradually from such a ferromagnetic tip, to obtain an up or down tip polarization. The applied external field to generate vortices on the sample is B=0.1T, which is smaller than the recovery field of the tip[22]. Thus, a ferromagnetic tip with out-of-plane polarization ( ↑ or ↓ ) can be used to detect the spin property of the ZBP. In Fig. 2B, we show the normalized ZBP spectra probed at the vortex center by using an Fe/W tip of M↑ and M↓ at B=0.1T(↑). Clearly, the ZBP in the parallel field-tip configuration (red) is higher than the ZBP in the anti-parallel configuration. To eliminate possible effect of spatial anisotropy to the spin-dependent tunneling conductance, we carried out the same measurements by reversing B from 0.1T(↑) to 0.1T(↓). As shown in the two right sub-panels in Fig. 2B, the height of the ZBP is found higher again in the parallel configuration. Since the normal tunneling contribution to the ZBP is essentially independent of the spin (see model calculation part, Fig. 3B), the spin dependent ZBP is in full agreement with the novel property of the MF to induce SSAR. We have measured the conductance far away from the center of the vortex, and found that the conductance is essentially independent of the spin polarization. Fig. 2C shows dI/dV at r=10 nm, which are independent of the spin. Therefore, the spin-dependence of the tunneling is a property at the vortex center, in further support of the scenario of the SSAR of the MF.

To make analyses more quantitative, we define spin polarization of the tunneling conductance,

$$P(B\uparrow) = \frac{G(B\uparrow,M\uparrow) - G(B\uparrow,M\downarrow)}{G(B\uparrow,M\uparrow) + G(B\uparrow,M\downarrow)} \qquad (2a)$$

and

$$P(B\downarrow) = \frac{G(B\downarrow,M\downarrow) - G(B\downarrow,M\uparrow)}{G(B\downarrow,M\downarrow) + G(B\downarrow,M\uparrow)}, \qquad (2b)$$

where $G(B\uparrow,M\uparrow) = dI(B\uparrow,M\uparrow)/dV$ is a spin polarization dependent conductance. The values of $P(B\uparrow)$ and $P(B\downarrow)$ obtained for r=0 from the data in Fig. 2B are about 7%, which corresponds to a relative increase of about 14% for the parallel configurations. P(B) for r=10 nm is essentially zero. As we will discuss below, P(B) is observed to vanish in conventional SC.

To model the 5QL $Bi_2Te_3$/$NbSe_2$ hetero-structure, we consider the surface of a 3D TI described by a Rashba spin-orbit coupled system[23,24], $H_0 = \frac{\alpha}{\hbar}(\vec{\sigma} \times \vec{p}) \cdot \hat{z}$ with $\hat{z}$ the normal direction to the surface, and its superconductivity is induced through the proximity to an s-wave SC. We assume a single vortex is created by an external magnetic field, when Zeeman splitting effect is much smaller than the orbital effect and can be neglected. Thus a MF is found inside the vortex core by solving Bogoliubov-de-Gennes equations. The wave function of such a MF is illustrated in Fig. 3A along with the first low lying excited quasi-particle state. The spin distribution of such low lying states are determined by the Rashba coupling together with superconducting order parameter. The fully spin polarization at the center of the vortex core is attributed to the fact that the Rashba coupling is caused by broken mirror symmetry at the surface and the rotation around the normal direction to the surface is still invariant.

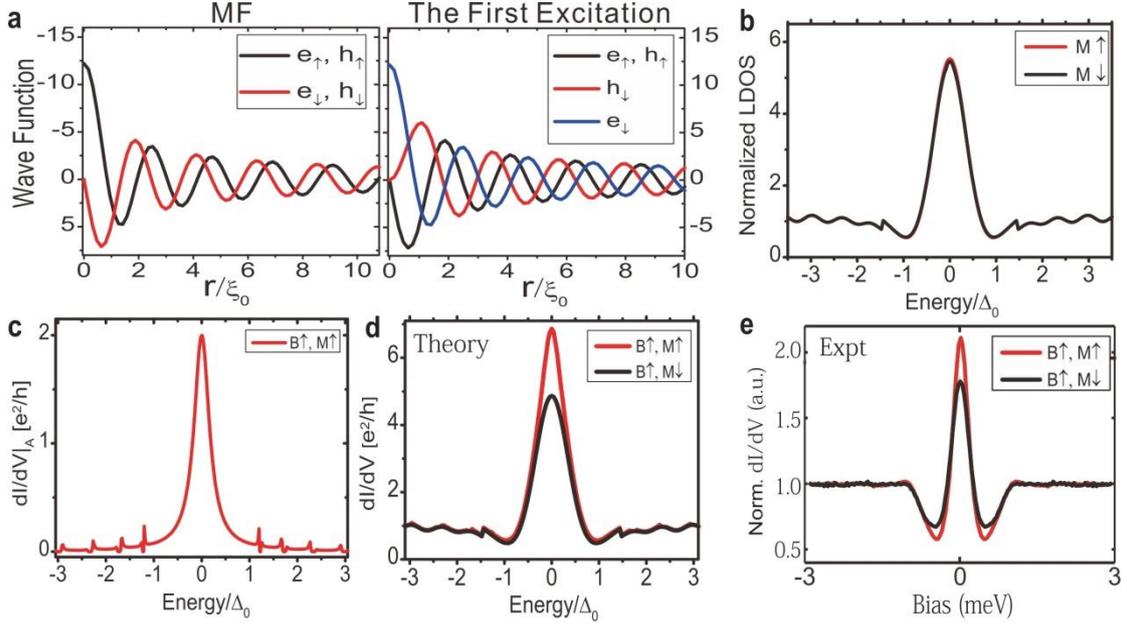

**Figure 3 | Model calculations for spin selective AR at the center of the vortex core in 2D topological superconductor. a.** Wavefunction of MF (left) and the first low lying quasi-particle state (right) inside a vortex core with magnetic field B upward. Spin components (↑,↓) of electron (e) and hole (h) are indicated in the figure. **b.** Normalized local density of states for spin ↑ (red) and spin ↓ (black) $at\ r = 0$, which are not very different. **c.** Calculated AR conductance at r=0 for the tip spin polarization M parallel to B. The AR conductance is zero for M anti-parallel to B. **d.** Spin dependent tunneling conductance in Eq. (1) estimated in a transparent limit. Red (black) curve is for M parallel (anti-parallel) to B. **e.** The same as in **d**, for experimentally measured conductance.

The Fe coated STM tip is modeled by a spin polarized metallic lead, which couples to the TI surface through a point contact at r=0. The junction setup is very similar to that used by He et al.[18], except that we use a 2D Rashba system instead of 1D quantum wire. We found that the AR contribution to the conductance, $dI(B\uparrow, M\uparrow)/dV|_A = 2e^2/h$, as plotted in Fig. 3B, while $dI(B\uparrow, M\downarrow)/dV|_A$ vanishes, which are the same as those obtained in the 1D wire[18]. Note that the superconducting order parameter is zero at the vortex center r=0, so that the only AR is via MF. The total tunneling conductance has a normal tunneling part, or the first term in Eq. (1). The relative weight between the two terms depends on the tunneling barrier. To make a comparison with experiment, we consider the limit of transparent barrier to estimate $dI(r, E)/dV|_n \sim 0.88 e^2/h$ for a given spin in the normal state, and obtain $dI(r = 0, E = 0)/dV|_n = dI(r, E)/dV|_n \, N(r = 0, E = 0)/N(r, E)$ with $N(r, E)$ the local density of states for the given spin [25]. The results for the total tunneling conductance[26-28] in Eq. (1) are plotted in Fig. 3C. We estimate $P(B\uparrow(\downarrow)) = 17\%$, which is 2.4 times of the experimental value of 7%. Fig. 3E is a plot of measured spin dependent dI/dV in the STM/STS, which compares with Fig. 3D of the theory quite well.

To further investigate the origin of the spin polarization dependence of the conductance, we have carried out a series of SP-STS measurements of vortex core state on conventional SCs, 3QL $Bi_2Te_3$ on top of $NbSe_2$ and bare $NbSe_2$ samples for comparison, where we do not expect MF[15]. The results are shown in Fig. 4A and 4B, in which four panels represent four external field-tip polarization configurations. As we can see clearly, the STS in 3QL $Bi_2Te_3$ and $NbSe_2$ show little difference in the ZBP strength between parallel and anti-parallel field-tip configuration for either positive field 0.1T ($\uparrow$) or the negative one. The values of $|P(B\uparrow(\downarrow))|$ for 3QL $Bi_2Te_3$ and $NbSe_2$ samples are smaller than 1%. This is in contrary to the spin dependent ZBP observed in 5QL $Bi_2Te_3/NbSe_2$.

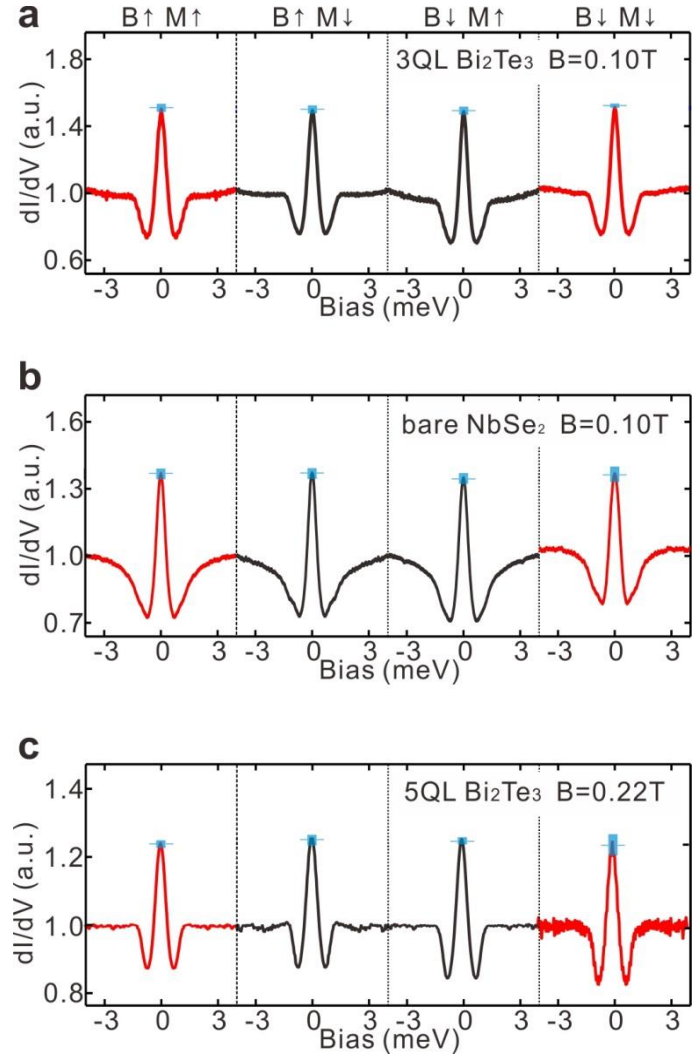

**Figure 4 | dI/dV curves at the center of a vortex core measured with a fully spin polarized tip.** Red colored curves are for polarization M parallel to B, and black curves are for M anti-parallel to B. **a.** and **b.** for 3QL $Bi_2Te_3$ on $NbSe_2$ and bare $NbSe_2$. In the measurements, B=0.1 T. **c.** for 5QL $Bi_2Te_3$ on $NbSe_2$ at B=0.22 T. The measurement temperature T=30 mK for all curves. The blue lateral lines give the average values of the intensities in multi measurements, the vertical bars are the standard error bars.

As the external magnetic field increases, the distance between vortex decreases, the interaction between vortices becomes stronger, which may destroy the MF inside vortices[15]. The SP STS measurements were also done at large magnetic fields for 5QL $Bi_2Te_3/NbSe_2$ sample and the results are shown in Fig.4C. From which, we can see that the spin-selective tunneling effects disappear when the magnetic field is larger than 0.22 T. All these results demonstrate that the MFs exist inside the vortices of 5QL $Bi_2Te_3$/$NbSe_2$ sample at a field around 0.1T, and spin-selective tunneling effects can be used to detect MFs.

The SSAR was observed at the center of the vortex core on 5QL $Bi_2Te_3$ films grown on $NbSe_2$ at 0.1 T. In this non-magnetic system, only MF can induce the SSAR. Together with non-selective signal obtained in other comparison systems, our work gave the definitive evidences of MF. It is also suggested that SSAR can be used in detection of MFs in 3D TI/SC heterostructures and other systems that host MFs with related spin-resolved techniques, MFs can be manipulated by the interaction between vortices. In addition, the spin current from MFs can be potentially used for spintronics.

**References:**
1    Majorana, E. Teoria simmetrica dell'elettrone e del positrone. *Nuovo Cim* **14**, 171-184 (1937).
2    Wilczek, F. Majorana returns. *Nat. Phys.* **5**, 614-618 (2009).
3    Elliott, S. R. & Franz, M. Majorana fermions in nuclear, particle, and solid-state physics. *Reviews of Modern Physics* **87**, 137-163 (2015).
4    Nayak, C., Simon, S. H., Stern, A., Freedman, M. & Das Sarma, S. Non-Abelian anyons and topological quantum computation. *Reviews of Modern Physics* **80**, 1083-1159 (2008).
5    Das Sarma, S., Nayak, C. & Tewari, S. Proposal to stabilize and detect half-quantum vortices in strontium ruthenate thin films: Non-Abelian braiding statistics of vortices in a $p_x+ip_y$ superconductor. *Phys. Rev. B* **73**, 220502 (2006).
6    Moore, G. & Read, N. Nonabelions in the fractional quantum hall effect. *Nuclear Physics B* **360**, 362-396 (1991).
7    Fu, L. & Kane, C. L. Superconducting Proximity Effect and Majorana Fermions at the Surface of a Topological Insulator. *Phys. Rev. Lett.* **100**, 096407 (2008).
8    Mourik, V. *et al.* Signatures of Majorana Fermions in Hybrid Superconductor-Semiconductor Nanowire Devices. *Science* **336**, 1003-1007 (2012).
9    Deng, M. T. *et al.* Anomalous Zero-Bias Conductance Peak in a Nb–InSb Nanowire–Nb Hybrid

|    | Device. *Nano Letters* **12** (2012). |
|----|---|
| 10 | Finck, A. D. K., Van Harlingen, D. J., Mohseni, P. K., Jung, K. & Li, X. Anomalous Modulation of a Zero-Bias Peak in a Hybrid Nanowire-Superconductor Device. *Phys. Rev. Lett.* **110**, 126406 (2013). |
| 11 | Lee, E. J. H. *et al.* Spin-resolved Andreev levels and parity crossings in hybrid superconductor-semiconductor nanostructures. *Nat. Nano.* **9**, 79-84 (2014). |
| 12 | Nadj-Perge, S. *et al.* Observation of Majorana fermions in ferromagnetic atomic chains on a superconductor. *Science* **346**, 602-607 (2014). |
| 13 | Wang, M.-X. *et al.* The Coexistence of Superconductivity and Topological Order in the $Bi_2Se_3$ Thin Films. *Science* **336** (2012). |
| 14 | Xu, J.-P. *et al.* Artificial Topological Superconductor by the Proximity Effect. *Phys. Rev. Lett.* **112**, 217001 (2014). |
| 15 | Xu, J.-P. *et al.* Experimental Detection of a Majorana Mode in the core of a Magnetic Vortex inside a Topological Insulator-Superconductor $Bi_2Te_3$/$NbSe_2$ Heterostructure. *Phys. Rev. Lett.* **114**, 017001 (2015). |
| 16 | Chiu, C.-K., Gilbert, Matthew J. and Hughes, Taylor L. Vortex lines in topological insulator-superconductor heterostructures. *Phys. Rev. B* **84**, 144507 (2011) |
| 17 | Xu, S.-Y. *et al.* Momentum-space imaging of Cooper pairing in a half-Dirac-gas topological superconductor. *Nat. Phys.* **10**, 943 (2014). |
| 18 | He, J. J., Ng, T. K., Lee, P. A. & Law, K. T. Selective Equal-Spin Andreev Reflections Induced by Majorana Fermions. *Phys. Rev. Lett.* **112**, 037001 (2014). |
| 19 | Haim, A., Berg, E., von Oppen, F. & Oreg, Y. Signatures of Majorana Zero Modes in Spin-Resolved Current Correlations. *Phys. Rev. Lett.* **114**, 166406 (2015). |
| 20 | Kawakami, T. & Hu, X. Evolution of Density of States and a Spin-Resolved Checkerboard-Type Pattern Associated with the Majorana Bound State. *Phys. Rev. Lett.* **115**, 177001 (2015). |
| 21 | Li, Z.-Z., Zhang, F.-C. & Wang, Q.-H. Majorana modes in a topological insulator/s-wave superconductor heterostructure. *Scientific Reports* **4** (2014). |
| 22 | Wiesendanger, R. Spin mapping at the nanoscale and atomic scale. *Reviews of Modern Physics* **81**, 1495-1550 (2009). |
| 23 | Sau, J. D., Tewari, S., Lutchyn, R. M., Stanescu, T. D. & Das Sarma, S. Non-Abelian quantum order in spin-orbit-coupled semiconductors: Search for topological Majorana particles in solid-state systems. *Phys. Rev. B* **82**, 214509 (2010). |
| 24 | Sau, J. D., Lutchyn, R. M., Tewari, S. & Das Sarma, S. Robustness of Majorana fermions in proximity-induced superconductors. *Phys. Rev. B* **82**, 094522 (2010). |
| 25 | Gygi, F. & Schlüter, M. Self-consistent electronic structure of a vortex line in a type-II superconductor. *Phys. Rev. B* **43**, 7609-7621 (1991). |
| 26 | Qing-feng Sun, Jian Wang, and Tsung-han Lin, Resonant Andreev reflection in a normal-metal–quantum-dot–superconductor system, *Phys. Rev. B* **59**, 3831 (1999). |
| 27 | Da Wang, Yuan Wan, and Qiang-Hua Wang, Model for Determining the Pairing Symmetry and Relative Sign of the Energy Gap of Iron-Arsenide Superconductors using Tunneling Spectroscopy. *Phys. Rev. Lett.* **102**, 197004 (2009). |
| 28 | Yang, K.-Y., Huang, K., Chen, W.-Q., Rice, T. M. & Zhang, F.-C. Andreev and Single-Particle Tunneling Spectra of Underdoped Cuprate Superconductors. *Phys. Rev. Lett.* **105**, 167004 (2010). |

**Acknowledgement** We thank Chui-Zhen Chen, Wei-Qiang Chen, Jun. He, K. T. Law, Qiang-Hua Wang, Dong-Hui Xu, for stimulating discussions. We wish to thank the Ministry of Science and Technology of China (2013CB921902, 2014CB921201, 2014CB921203, 2014CB921103, 20130073120081, 2012CB927401), NSFC (11521404, 11227404, 11274269, 11374256, 11374140, 11504230，11574202，11134008, 11374206, 11274228, 11574201) for partial support.

**Author contributions** HHS and KWZ conducted the experiments with the help of GYW and HYM. SCL provides all the supports for the experiments and supervised KWZ. LHH and CL carried out the calculations with the help from YZ, LF and FCZ. JFJ designed the experiments with the help from FCZ. ZAX provided NbSe2 crystals. SCL, CLG, DDG, CHL, DQ, LF, FCZ and JFJ analyzed the data. HHS, FCZ and JFJ wrote most of the paper. All the authors discussed the results.

**Author Information** Reprints and permissions information is available at www.nature.com/reprints. The authors declare no competing financial interests. Readers are welcome to comment on the online version of the paper. Correspondence and requests for materials should be addressed to J.F.J. (jfjia@sjtu.edu.cn), F.C.Z. (fuchun@hku.hk) or S.C.L. (scli@nju.edu.cn).

## METHODS

**Experiment details.** The experiments were performed in an ultrahigh vacuum (UHV) low temperature STM-MBE joint system (Unisoku dilution LT UHV STM with SC Magnet USM1600). The *2H*-NbSe$_2$ substrate was synthesized by chemical vapor transport (CVT) method and *in situ* cleaved in UHV with base pressure $1 \times 10^{-10}$ Torr. Bi$_2$Te$_3$ film was then grown on the substrate at 500K using standard Knudsen cell sources. STM/STS measurements were taken at a nominal temperature of 30 mK with $^3$He/$^4$He dilution refrigerator. Electrochemically etched W tips were adopted for normal topography and spectroscopy measurements, and Fe-coated W tips spin polarization measurements. dI/dV spectra were taken via lock-in technique with a modulation of 0.1 mV at frequency of 991 Hz, and a set point of 0.2 nA. dI/dV mapping of vortices were measured at zero bias with a set point of 0.1 nA and feedback loop off. All STS curves were normalized to the dI/dV value of the spectra at bias energy outside the superconducting gap, so that the intensities of the ZBP under the four field-tip configurations can be compared to each other.

**Calculation details.** In the calculations we set superconducting gap $\Delta_0 = 1$ (1 meV in experiment) as the energy unit, the size of the vortex $\xi_0$ (35 nm in experiment) as the length unit defined as the distance from the center at which the superconducting gap $\Delta(\xi_0) = \tanh(1)\Delta_0$, and $\alpha = 35$, and Fermi energy $E_F = 100$. The obtained first excited state energy is 0.04. We use energy broadening width $\eta = 0.4$, and the hopping between the lead and the vortex core center $t_c = 0.2$.